\documentclass[
 reprint,
 amsmath,amssymb,
 aps,
prl,
]{revtex4-2}

\usepackage{graphicx}
\usepackage{dcolumn}
\usepackage{bm}
\usepackage[hidelinks]{hyperref}

\newcommand{\beginsupplement}{%
  \clearpage
  \onecolumngrid
  \setcounter{page}{1}

  \setcounter{figure}{0}
  \setcounter{table}{0}
  \setcounter{equation}{0}
  \renewcommand{\thefigure}{S\arabic{figure}}
  \renewcommand{\thetable}{S\arabic{table}}
  \renewcommand{\theequation}{S\arabic{equation}}
}

\newcommand{\suppsection}[1]{%
  \section{#1}%
  \addcontentsline{sptoc}{section}{\protect\numberline{\thesection}#1}%
}

\begin{document}

\title{Low-Dimensional Reduction Theory for Populations of Globally Coupled Phase Oscillators with Multiharmonic Coupling: A Method Based on OPUC Theory}

\author{Kai Tokunaga}
\email{tokunaga.kai25a@c.k.u-tokyo.ac.jp}
\affiliation{
 Department of Complexity Science and Engineering, The University of Tokyo, Kashiwa, Chiba 277-8561, Japan
}

\date{\today}

\begin{abstract}
Low-dimensional reduction theories such as the Ott--Antonsen ansatz have played a crucial role in the study of populations of coupled oscillators. However, most of these theories apply only to models in which the interaction is described by a single harmonic component, limiting their use in more realistic oscillator models. Using the theory of orthogonal polynomials on the unit circle (OPUC), we develop a low-dimensional reduction theory for populations of globally coupled phase oscillators with multiharmonic coupling. We show theoretically and numerically that it is exact for uniformly rotating solutions and provides a good approximation for nonequilibrium solutions.
\end{abstract}

\maketitle

Populations of coupled limit-cycle oscillators have long attracted attention because of their wide-ranging applications in fields such as biology, chemistry, engineering, and physics \cite{kuramoto1984chemical,winfree2001,Pikovsky_Rosenblum_Kurths_2001,strogatz2003sync}. An important theoretical framework for studying populations of coupled limit-cycle oscillators is phase reduction \cite{kuramoto1984chemical,winfree2001,kuramoto1975,WINFREE1967}. Through phase reduction, the original complex system is reduced to a population of phase oscillators described solely by phase variables. However, even populations of phase oscillators remain highly complex when the number of oscillators is very large, including the case of infinitely many oscillators. In such cases, low-dimensional reduction theories \cite{Ott2008,WATANABE1994,Tyulkina2018,Bick2020,Cestnik2022,Cestnik2022feb}, such as the Ott–Antonsen ansatz \cite{Ott2008}, provide useful theoretical tools. These theories reduce populations of phase oscillators to low-dimensional dynamical systems described by macroscopic
variables.

However, most low-dimensional reduction theories have been applicable only when the interaction can be described by a single harmonic. Since coupling functions obtained through phase reduction contain multiple harmonics in many cases, conventional low-dimensional reduction theories have not been sufficient to capture complex interactions in realistic populations of limit-cycle oscillators. Indeed, multiharmonic coupling is known to play an important role in diverse phenomena such as collective chaos \cite{Leon2022,Clusella2021}, heteroclinic cycles \cite{Hansel1993,Kori2001}, and multi-branch entrainment \cite{Daido1996Aug,Komarov2013}. However, because of the limitations of existing low-dimensional reduction theories, research in this field has largely focused on populations of phase oscillators with single-harmonic coupling. In this paper, using a method based on the theory of orthogonal polynomials on the unit circle (OPUC) \cite{simon20051,szego1939}, we develop an approximate low-dimensional reduction theory for populations of globally coupled phase oscillators with multiharmonic coupling.

The results of this paper can be regarded as an extension of the previous results \cite{Tonjes2020} on uniformly rotating solutions in populations of globally coupled phase oscillators with multiharmonic coupling subject to independent Cauchy noise. We first show that the probability measure providing a low-dimensional description of the uniformly rotating solutions found in \cite{Tonjes2020} is in fact the Bernstein–Szegő measure \cite{simon20051} obtained through truncation of the Verblunsky coefficients. We then show that, for nonequilibrium solutions other than uniformly rotating ones, the higher-order Verblunsky coefficients—which vanish for uniformly rotating solutions—remain sufficiently small under a quasistationary approximation. As a result, the low-dimensional reduction based on truncation of the Verblunsky coefficients provides a highly accurate approximation to the original infinite-dimensional dynamics.

In this paper, we consider a population of globally coupled identical phase oscillators with harmonics up to the $L$-th order, subject to independent normalized Cauchy noise with noise intensity $\gamma$, as follows:
\begin{equation}
\label{eq:phase oscillators}
\dot{\theta}_i = \omega + \sum_{l=1}^{L}[h_l(t)e^{il{\theta}_i}+\overline{h_l(t)}e^{-il{\theta}_i}] + \gamma \xi_i(t)
\end{equation}
We also treat the thermodynamic limit of an infinite population. In this case, the phase density function $\rho(\theta, t)$ obeys the following continuity equation:
\begin{equation}
\label{eq:continuity equation}
\frac{\partial \rho(\theta, t)}{\partial t} + \frac{\partial}{\partial \theta} (h(\theta,t)\rho(\theta, t) ) = \gamma \left|\frac{\partial}{\partial \theta}\right|\rho(\theta, t)
\end{equation}
Here, the right-hand side of \eqref{eq:continuity equation} represents the effect of the Cauchy noise, and $\left|\frac{\partial}{\partial \theta}\right|$ is an operator that acts on each term in the Fourier expansion of a periodic function as
$\left|\frac{\partial}{\partial \theta}\right|e^{i n\theta}=-\left|n\right|e^{i n\theta}$.
We also define
$h(\theta,t)=\omega + \sum_{l=1}^{L}[h_l(t)e^{il\theta}+\overline{h_l(t)}e^{-il\theta}]$.
We define the moments of the probability measure on the unit circle determined by the phase density function $\rho(\theta, t)$ as $Z_n(t) = \int_0^{2\pi} e^{-in\theta} \rho(\theta, t)\, d\theta$. The commonly used Kuramoto--Daido order parameter \cite{DAIDO1996} is simply the complex conjugate of this moment. In this paper, we consider a Kuramoto--Sakaguchi-type coupling function \cite{sakaguchi1986}. Therefore, letting $\varepsilon_l$ be the coupling strength and $\tau_l$ be the phase shift, we write $h_l(t)=-\frac{\varepsilon_l e^{i\tau_l}}{2i}Z_l(t)$. The phase density function can be expanded in terms of the moments as $\rho(\theta, t)= \frac{1}{2\pi}\left\{1+\sum_{n=1}^{\infty} \bigl[Z_n(t) e^{in\theta}+\overline{Z_n(t)}e^{-in\theta}\bigr]\right\}$. Substituting this into  \eqref{eq:continuity equation}, we obtain the following infinite set of ODEs for the moments:
\begin{equation}
\label{eq:moment ODE}
 \dot{Z}_n = (-in\omega-\gamma|n|) Z_n - in \sum_{l=1}^{L}\bigl[h_l Z_{n-l} +  \overline{h_l} Z_{n+l}\bigr]
\end{equation}
Here, when $L=1$, it is known that for oscillators subject to independent Cauchy noise, the Ott--Antonsen ansatz $Z_n = Z_1^n$ yields the following finite-dimensional ODE closed in $Z_1$ \cite{Tonjes2020,Tanaka2020,Cestnik2022}:
\begin{equation}
\label{eq:OA equation}
 \dot{Z}_1 = (-i\omega-\gamma) Z_1 - i \bigl[h_1 +  \overline{h_1} Z_{1}^2\bigr].
\end{equation}
Moreover, the Ott--Antonsen ansatz has also been applied to the case where the coupling consists of a single higher-order harmonic component \cite{Skardal2011}. The aim of this study is to derive a useful finite-dimensional reduced ODE for the multiharmonic case with $L\geq 2$.

Assume that the probability measure $\rho(\theta,t)d\theta$ determined by the phase density function $\rho(\theta,t)$ is nontrivial (thus excluding complete synchronization or complete clustering, for which the absolute value of the moment is exactly equal to $1$). We define the inner product on the Hilbert space $L^2(\partial\mathbb{D},\rho(\theta,t)\,d\theta)$ by $\langle f,g \rangle = \int_0^{2\pi}  \overline{f(\theta)}\,g(\theta) \, \rho(\theta,t)\,d\theta$. Since $\{1,z,z^2,z^3,\ldots\}$ is linearly independent in $L^2(\partial\mathbb{D},\rho(\theta,t)\,d\theta)$, we can define the monic orthogonal polynomials $\Phi_n(z)$ \cite{simon20051,szego1939} by Gram--Schmidt orthogonalization. The resulting $\Phi_n(z)$ is uniquely characterized as the monic polynomial of degree $n$ that is orthogonal to $\{1,\ldots,z^{n-1}\}$. Moreover, for the reversed polynomial of the monic orthogonal polynomial $\Phi_n(z)$, defined by $\Phi_n^*(z):=z^n \overline{\Phi_n(1/\overline{z})}$, the polynomial $\Phi_n^*(z)$ is uniquely characterized as a polynomial of degree at most $n$ with constant term $1$ that is orthogonal to $\{z,z^2,\ldots,z^{n}\}$. Then $\Phi_n(z)$ and $\Phi_n^*(z)$ satisfy the following Szeg\H{o} recurrences \cite{simon20051,szego1939}:
\begin{equation}
\label{eq:Szego recurrence1}
\Phi_{n+1}(z) = z\Phi_n(z) - \overline{\alpha_n}\Phi_n^*(z)
\end{equation}
\begin{equation}
\label{eq:Szego recurrence2}
\Phi_{n+1}^*(z) = \Phi_n^*(z) - \alpha_n z\Phi_n(z)
\end{equation}
The coefficients $\alpha_n$ appearing here are called the Verblunsky coefficients. The Verblunsky coefficients have a remarkable property: for any sequence $\{\alpha_n\}_{n=0}^{\infty}$ satisfying $|\alpha_n|<1$, there exists a unique nontrivial probability measure on the unit circle whose Verblunsky coefficients are $\{\alpha_n\}_{n=0}^{\infty}$. This is known as Verblunsky's theorem \cite{Verblunsky1935}. By Verblunsky's theorem, if the Verblunsky coefficient sequence $\{\alpha_n\}_{n=0}^{\infty}$ corresponding to a certain probability measure is given, then for any fixed natural number $N$, the sequence obtained by setting $\alpha_n = 0$ for all $n \ge N$ again determines a probability measure on the unit circle. The measure on the unit circle defined in this way is called the ($N$-th order) Bernstein--Szeg\H{o} measure \cite{simon20051}, and it provides an approximation of the original probability measure by a probability measure described by only finitely many parameters. In this paper, we show that the Bernstein--Szeg\H{o} measure plays an important role in the low-dimensional reduction for populations of  globally coupled phase oscillators. As a first simple correspondence that can be observed, the density function of the first-order Bernstein--Szeg\H{o} measure, $\rho(\theta)=\frac{1}{2\pi}\frac{1-|\alpha_0|^2}{|1-e^{i\theta}\alpha_0|^2}$, is seen to coincide with the low-dimensional manifold represented by the Ott--Antonsen ansatz (the OA manifold), where $\alpha_0=Z_1$.

In this study, we first consider the case of uniformly rotating solutions. When we consider the system in a rotating frame, taking $\theta \to \theta-\Omega t$, with $\Omega$ denoting the frequency of uniform rotation, the solution can be regarded as a stationary solution, and from \eqref{eq:moment ODE} we obtain the following homogeneous linear recurrence relation of order $2L$ with constant coefficients for $n\ge 1$:
\begin{equation}
\label{eq:stationary recurrence relation}
  (\omega-\Omega - i\gamma) Z^{st}_n
  + \sum_{l=1}^{L}\bigl[\,h^{st}_l Z^{st}_{n-l} + \overline{h^{st}_l} Z^{st}_{n+l}\bigr] = 0 \quad(n\ge 1)
\end{equation}
Since the characteristic equation of this recurrence relation can be written as $
  \sum_{l=1}^{L}\bigl[\,h^{st}_l \lambda^{-l}
  + \overline{h^{st}_l} \lambda^{l}\bigr] = -(\omega-\Omega - i \gamma)
$, Rouch\'e's theorem implies that it has $L$ roots inside the unit disk and $L$ roots outside the unit disk. Here, since $|Z_n| < 1$, it follows that the coefficients associated with the roots outside the unit disk must be zero. Moreover, since cases in which multiple roots occur are expected to be exceptional, we may write $Z^{st}_n = \sum_{m=1}^{L}C_m \lambda_m^{n}$ using the roots inside the unit disk. Note that, because this recurrence relation contains moments down to $Z_{1-L}$, this solution is valid for $n \ge 1-L$. Furthermore, in order to satisfy consistency as moments, we must have $Z_0 = 1$ and $Z_{-m} = \overline{Z_m}$ for $m = 1,\ldots,L-1$. Together with the condition that the coefficients of the roots outside the unit disk are set to $0$, these determine $2L$ initial conditions, and hence the solution of the recurrence relation \eqref{eq:stationary recurrence relation} is uniquely determined. The discussion up to this point is based on \cite{Tonjes2020}.

Taking into account that, above, all coefficients corresponding to the roots outside the unit disk were set equal to $0$, we see that \eqref{eq:stationary recurrence relation} is in fact equivalent to the following $L$-th order recurrence relation determined by the characteristic polynomial $P(z) = \prod_{m=1}^L (z - \lambda_m)$:
\begin{equation}
\label{eq:Lth stationary recurrence relation}
  Z^{st}_{n} - e_1(\lambda) Z^{st}_{n-1} + \cdots + (-1)^L e_L(\lambda) Z^{st}_{n-L} = 0
  \quad  (n \ge 1)
\end{equation}
Here, $e_k(\lambda)$ denotes the $k$-th elementary symmetric polynomial in $\{\lambda_1,\ldots,\lambda_L\}$. Since the previous solution holds for $n \ge 1-L$, \eqref{eq:Lth stationary recurrence relation} also holds for $n \ge 1$. Writing $P(z)=\sum_{m=0}^{L} d_{L-m} z^m$, the recurrence relation \eqref{eq:Lth stationary recurrence relation} can be written, with respect to the inner product on $L^2(\partial\mathbb{D},\rho(\theta,t)\,d\theta)$, as
\begin{equation}
\label{eq:recurrence relation in inner product}
  \langle z^n,\sum_{m=0}^{L}d_{m} z^m \rangle = 0 \quad  (n \ge 1)
\end{equation}
Therefore, since $\sum_{m=0}^{L}d_{m} z^m$ is a polynomial of degree $L$ with constant term $1$ that is orthogonal to $\{z,\ldots,z^{L}\}$, we have $\sum_{m=0}^{L}d_{m} z^m= \Phi_L^{*}(z)$. Furthermore, since \eqref{eq:recurrence relation in inner product} also holds for $n \ge L$, it follows that $\Phi_n^{*}(z) = \Phi_L^{*}(z)$ for $n \ge L$. Hence, by the Szeg\H{o} recurrence \eqref{eq:Szego recurrence2}, we have $\alpha_n = 0$ for $n \ge L$, and therefore the uniformly rotating solution of this model is described by the $L$-th order Bernstein--Szeg\H{o} measure. Moreover, for the monic orthogonal polynomial, we find that $\Phi^{st}_L(z) = \sum_{m=0}^{L} \overline{d_{L-m}} z^m=\overline{P(\overline{z})}$.

Next, we consider nonequilibrium solutions. Taking the rotating frame $\theta \to \theta-\Omega t$, where $\Omega$ is the mean frequency of the nonequilibrium solution, and considering a quasistationary approximation assuming the time variation of the nonequilibrium solution is sufficiently slow, from \eqref{eq:moment ODE} we obtain
\begin{equation}
\label{eq:quasistationary recurrence relation}
  (\omega-\Omega - i\gamma) Z_n
  + \sum_{l=1}^{L}\bigl[\,h_l Z_{n-l} + \overline{h_l} Z_{n+l}\bigr] = 0 \quad(n\ge 1)
\end{equation}
Since \eqref{eq:quasistationary recurrence relation} satisfies a recurrence relation of the same form as that for the uniformly rotating solution, the moments $Z_n^{qst}$  satisfying \eqref{eq:quasistationary recurrence relation} can be written as $Z_n^{qst}=\sum_{m=1}^{L}C_m\lambda_m^{n}$ in terms of the roots inside the unit disk. It follows that the solution is described by the $L$-th order Bernstein--Szeg\H{o} measure. Introducing the slow time $\tau=\varepsilon t$, we expand $Z_n=Z_n^{qst}+\varepsilon u_n+O(\varepsilon^2)$. Then, at $O(\varepsilon)$, we obtain the following inhomogeneous linear recurrence relation with constant coefficients:
\begin{equation}
\label{eq:inhomogeneous recurrence relation}
\begin{split}
&i\sum_{m=1}^{L} A_m\lambda_m^{n}+i\sum_{m=1}^{L} \frac{dC_m}{d\tau}\frac{\lambda_m^{n}}{n} =\\
&(\omega-\Omega - i\gamma) u_n
  + \sum_{l=1}^{L}\bigl[\,h^{qst}_l u_{n-l}+ \overline{h^{qst}_l}u_{n+l}\bigr] \quad  (n \ge 1)
\end{split}
\end{equation}
Here, $A_m$ denotes the coefficient of the inhomogeneous term of the form $\lambda_m^n$. The solution of \eqref{eq:inhomogeneous recurrence relation} therefore contains large resonant terms arising from the inhomogeneous terms of the forms \(\lambda_m^n\) and \(\lambda_m^n/n\). Therefore, we consider a coordinate transformation $Z_n=\widetilde{Z}_n+\varepsilon\sum_{m=1}^{L} G_{m,n}\lambda_m^{(n+L)}$ that cancels these inhomogeneous terms appropriately. It suffices to choose $G_{m,n}$ as follows:
\begin{equation}
\label{eq:Gmn}
G_{m,n}=i\frac{A_m}{\lambda_mQ'(\lambda_m)}(n+L)+i \frac{\frac{dC_m}{d\tau}}{\lambda_mQ'(\lambda_m)}H_{n+L}
\end{equation}
Here, $Q(z)$ denotes the characteristic polynomial of the homogeneous system in \eqref{eq:inhomogeneous recurrence relation}, and $H_n$ denotes the harmonic number. Letting $\widetilde{Z}_n=Z_n^{qst}+\varepsilon\widetilde{u}_n$, we consider the recurrence relation \eqref{eq:inhomogeneous recurrence relation} for $\widetilde{u}_n$. Then only inhomogeneous terms of order $\frac{\lambda_m^n}{n^2}$ or smaller remain, and the resonant terms are the same order as the homogeneous solution. Therefore, in the original coordinate system, letting $\lambda_{max}$ denote the root inside the unit disk with the largest modulus, we obtain $u_n= G_{max,n}\lambda_{max}^{n}+O(\lambda_{max}^n)$ for $n\ge 1-L$.

Next, using this result, we estimate the magnitude of the Verblunsky coefficients of order $L$ and higher. The Verblunsky coefficients of order $L$ and higher can be expanded in $\varepsilon$ as follows (see the Supplemental Material):
\begin{equation}
\label{eq:verblunsky eps expansion}
\alpha_n =  \frac{\varepsilon}{||\Phi_L^{*\,qst}||^2}
\sum_{j=0}^{L} \sum_{k=0}^{L} 
d_j\,d_k\, u_{(n+1)-(j+k)} +O(\varepsilon^2) \,  (n \ge L)
\end{equation}
Here, $d_m$ denotes the coefficient of the reversed polynomial of the monic orthogonal polynomial, $\Phi_L^{*\,qst}(z)=\sum_{m=0}^{L}d_{m} z^m$. Substituting $u_n$ into \eqref{eq:verblunsky eps expansion}, we find that terms such as $\lambda_m^n$ and $n\lambda_m^n$ vanish, since $1/\lambda_m$ is a zero of $\Phi_L^{*\,qst}(z)$ and \eqref{eq:verblunsky eps expansion} is a double sum.
Therefore, the terms involving $H_n$ become the dominant terms. However, since $ H_{n+L-(j+k)} = H_{n+L}-\frac{j+k}{n+L}+O(\frac{1}{n^2}) $ can be rewritten in this form, the first two terms become terms of the form $\lambda_m^n$ and $n\lambda_m^n$ in (13), and hence these also vanish. Therefore, it follows that $\alpha_n=O(\frac{\varepsilon\lambda_{max}^n}{n^2})$. Moreover, since this is an estimate at each time, if we denote the transient time by $T$ and set $\Lambda_{\max}=\max_{1\le m\le L,\ t>T}|\lambda_m(t)|$, then after the transient we can estimate $\alpha_n=O(\frac{\varepsilon\Lambda_{\max}^n}{n^2})$ uniformly in time. Furthermore, if the width of the attractor is $O(\delta)$, then $\frac{dC_m}{d\tau}$ is also expected to be $O(\delta)$, and hence we can estimate $\alpha_n=O(\frac{\varepsilon\delta\Lambda_{\max}^n}{n^2})$. This estimate suggests that, compared with the geometric decay determined by the largest root, the additional factor $\frac{1}{n^2}$ makes the decay even faster. Note that the constant involved depends on \(Q'(\lambda_{\max})\),
so when characteristic roots approach one another, the constant is expected
to become large. The quasistationary approximation may also remain valid
even when the nonequilibrium solution is not slowly varying. This is
expected when relaxation toward the quasistationary manifold is sufficiently
fast relative to this variation. Establishing a similar estimate for
\(\alpha_n\) in more general settings is left for future work.
\begin{figure}
\includegraphics{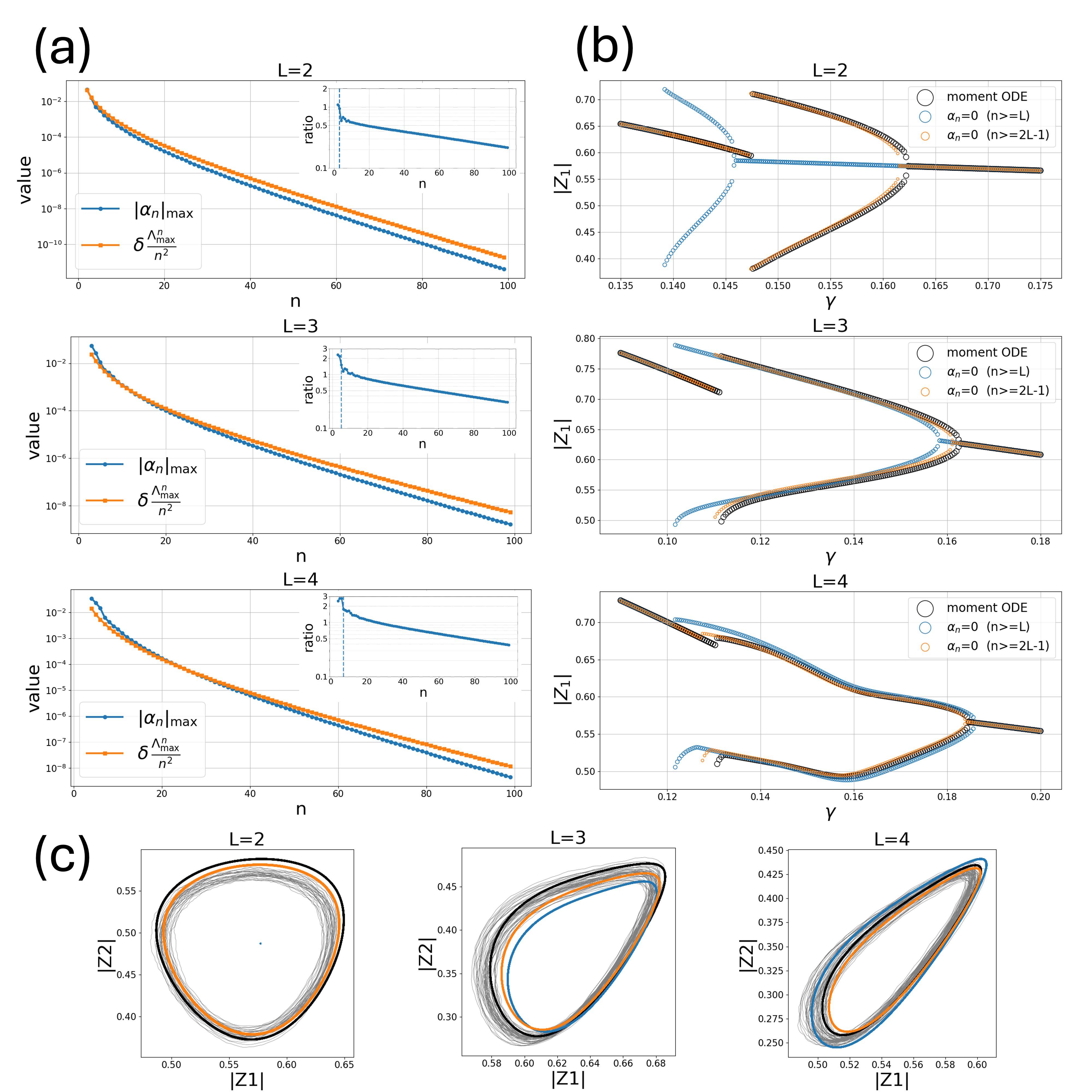}
\caption{\label{fig:epsart} (a) Order estimate for the Verblunsky coefficients, (b) maxima and minima of $|Z_1|$ for different $\gamma$, and (c) limit-cycle trajectories. In (c), black, blue, orange, and thin gray denote the moment ODEs, $N=L$ reduction, $N=2L-1$ reduction, and Langevin equation, respectively. In (a)–(c), parameters are: for $L=2$, $\epsilon_1=1.0,\epsilon_2=0.7,\tau_1=-0.2,\tau_2=-1.0$; for $L=3$, $\epsilon_1=1.0,\epsilon_2=0.7,\epsilon_3=0.5,\tau_1=-0.2,\tau_2=-1.0,\tau_3=-0.9$; and for $L=4$, $\epsilon_1=1.0,\epsilon_2=0.7,\epsilon_3=0.5,\epsilon_4=0.4,\tau_1=-0.2,\tau_2=-1.0,\tau_3=-0.9,\tau_4=-0.8$. Moreover, in (a) and (c), $\gamma=0.158$ for $L=2$, $\gamma=0.15$ for $L=3$, and $\gamma=0.165$ for $L=4$.}
\end{figure}

Figs.~1(a) and 2(a) numerically confirm this order estimate for the case of a periodic oscillation in the amplitude of the order parameter (breathing phenomena) and for the case of collective chaos, respectively. For the numerical computations, we simulate the moment ODE \eqref{eq:moment ODE} truncated to the first $1000$ moments using RK45. Then, from the resulting time evolution of the moments, we compute the time evolution of the Verblunsky coefficients using the Levinson--Durbin algorithm. We then compare the post-transient maximum absolute values $|\alpha_n|_{\max}$ of the first 100 Verblunsky coefficients with $\frac{\delta\Lambda_{\max}^n}{n^2}$. Here, letting $\langle \cdot \rangle$ denote the long-time average, we computed $\delta$ as $\delta=\max_{ t>T}\sqrt{\sum_{k=1}^{100}(|Z_k|-\langle |Z_k| \rangle)^2}$, and we did not take the timescale $\varepsilon$ into account. The corresponding ratio plots are also shown as insets. The numerical results show that this order estimate is particularly sharp for periodic solutions. For chaotic dynamics, it also provides a sharp estimate, but less sharp than in the periodic case. This is presumably because, in the chaotic case, values near the maximum are attained only on short time scales. 

This order estimate implies that, when the original system follows a nonequilibrium solution, the finite-dimensional system obtained by setting $\alpha_n=0 \ (n\ge N)$ for $N\ge L$ provides an approximation with accuracy $O(\frac{\varepsilon\delta\Lambda_{max}^N}{N^2})$ (see the Supplemental Material for the ODEs for the Verblunsky coefficients). This finite-dimensional system is a dynamical system on a finite-dimensional manifold represented by the $N$-th order Bernstein--Szeg\H{o} measure. Moreover, recalling the previous discussion of uniformly rotating solutions, we conclude that this finite-dimensional system provides a low-dimensional reduction that approximates the original system with accuracy $O(\frac{\varepsilon\delta\Lambda_{max}^N}{N^2})$ and exactly coincides with the original system for uniformly rotating solutions. Note that this low-dimensional reduction generalizes the OA ansatz for $N=L=1$ to the multiharmonic case with $L\ge 2$. Another important point is that, by Verblunsky's theorem, the phase distribution obtained by setting $\alpha_n=0 \ (n\ge N)$ is again a probability measure. For example, in the case $L=1$ with Gaussian noise, a low-dimensional reduction obtained by truncating the circular cumulants is known \cite{Tyulkina2018,Goldbin2019}; however, it is  known that the resulting low-dimensional phase distribution is not consistent as a probability measure \cite{Goldbin2019}.
\begin{figure}
  \includegraphics{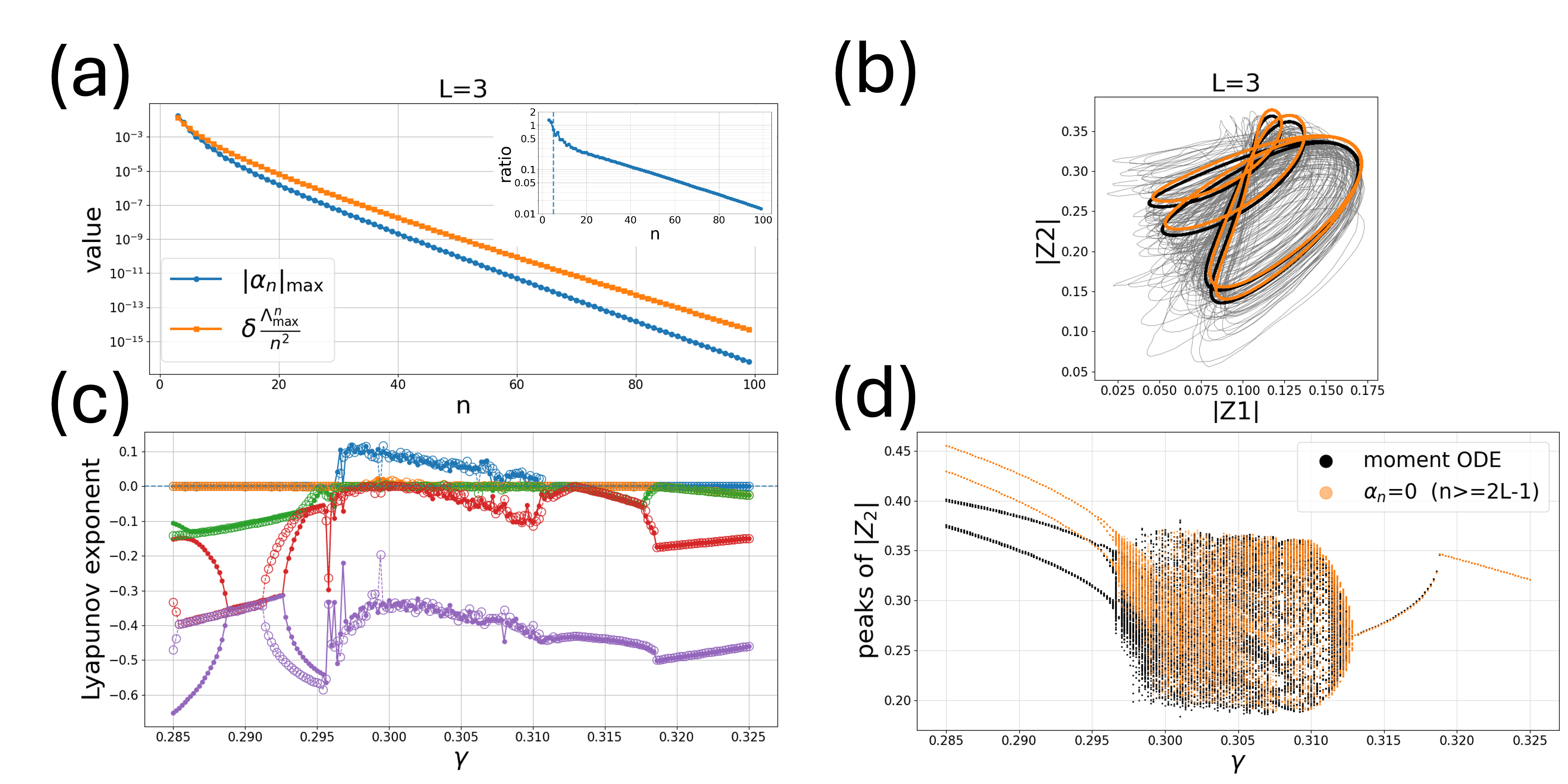}
  \caption{(a) Order estimate for the Verblunsky coefficients, (b) period-4 orbit, (c) five largest Lyapunov exponents for different $\gamma$, and (d) peak values of $|Z_2|$ for different $\gamma$. In (b), black, orange, and thin gray denote the moment ODEs, $N=2L-1$ reduction, and Langevin equation, respectively. In (c), filled circles/solid lines and open circles/dashed lines denote the moment ODEs and $N=2L-1$ reduction, respectively. In (a)–(d), $L=3$ and $\epsilon_1=3.0,\epsilon_2=2.0,\epsilon_3=2.0,\tau_1=-1.6,\tau_2=-1.2,\tau_3=-1.2$. In (a), $\gamma=0.3$, and in (b), $\gamma=0.296$.}
\end{figure}

From this point on, we compare the dynamics of the low-dimensional reduced system obtained by setting $\alpha_n=0\ (n\ge N)$ with the dynamics of the original system. Since the ODEs written in terms of the Verblunsky coefficients are complicated, we instead use an equivalent ODE system written in terms of the moments. When the measure is a Bernstein--Szeg\H{o} measure, the moments satisfy a linear recurrence relation with constant coefficients. Using this relation, higher-order moments can be expressed in terms of lower-order moments, thereby closing the ODE system. Setting $\Phi_N^*(z)=\sum_{m=0}^{N}d_{m}^{(N)} z^m$, it is sufficient to consider the following ODEs:
\begin{equation}
\label{eq:finite moment ODE}
\begin{split}
 &\dot{Z}_n = (-in\omega-\gamma|n|) Z_n \\
 &- in \sum_{l=1}^{L}\bigl[h_l Z_{n-l} +  \overline{h_l} Z_{n+l}\bigr] 
 \quad(1\le n\le N)
\end{split}
\end{equation}
\begin{equation}
\label{eq:Nth reccurence}
Z_n=-\sum_{m=1}^{N}d_{m}^{(N)} Z_{n-m} \quad(N+1\le n\le N+L)
\end{equation}
\begin{equation}
\label{eq:Toeplitz for dm}
 \begin{pmatrix}
Z_0   & Z_1     & \cdots & Z_{N-1} \\
Z_{-1}& Z_0     &  \cdots       & Z_{N-2} \\
\vdots& \vdots  &        & \vdots \\
Z_{-N+1}& Z_{-N+2}& \cdots & Z_0
\end{pmatrix} 
\begin{pmatrix}
d_N^{(N)} \\
\vdots\\
d_1^{(N)}
\end{pmatrix}=-
\begin{pmatrix}
Z_N \\
\vdots\\
Z_{1}
\end{pmatrix}
\end{equation}
Here, the coefficients $d_m^{(N)}$ can in practice be computed using the Levinson--Durbin algorithm. In the ratio plots in Figs.~1(a) and 2(a), $n=2L-1$ is indicated by a blue dotted line, and the values drop sharply beyond that point compared with those before it. Moreover, the Verblunsky coefficients with $n\ge2L-1$ are not directly driven by the first $L$ Verblunsky coefficients (see the Supplemental Material). Therefore, choosing $N=2L-1$ is considered to be a natural choice.

Figure~1(b) shows, for $L=2,3,4$, the numerically computed maximum and minimum values of $|Z_1|$ as the parameter $\gamma$ is decreased, for (i) the system obtained by truncating the moment ODE \eqref{eq:moment ODE}, (ii) the finite-dimensional system obtained with $N=L$, and (iii) the finite-dimensional system obtained with $N=2L-1$. Here, we consider the parameter range in which a periodic solution is born from the uniformly rotating solution through a Hopf bifurcation and disappears through a saddle-node-on-invariant-circle bifurcation. First, for the uniformly rotating solution, one can see that the two finite-dimensional systems capture the results of the original system exactly. On the other hand, for the periodic solution, the agreement is not perfect, but the accuracy is significantly improved in the case $N=2L-1$ compared with the case $N=L$. This tendency is particularly pronounced when $L=2$. Note that in Fig.~1(b), there is a parameter region where the uniformly rotating solution appears to exist only in the moment ODE; however, this region is actually bistable, and the finite-dimensional system also has a corresponding uniformly rotating solution. Fig.~1(c) further compares the limit-cycle trajectories obtained from (i), (ii), and (iii) with those obtained by simulation of the Langevin equation \eqref{eq:phase oscillators}, except that (ii) is a fixed point for $L=2$. Again, the accuracy improves for $N=2L-1$ compared with $N=L$. Fig.~2(b) compares the Langevin equation, the moment ODE, and the finite-dimensional system with $N=2L-1$ for the period-4 orbit that appears before the system reaches chaos via a period-doubling cascade. Figs.~2(c) and 2(d) compare, for different values of $\gamma$, the five largest Lyapunov exponents and the peaks of $|Z_2|$ between the moment ODEs and the finite-dimensional system with $N=2L-1$. These results show that the system obtained by truncating the Verblunsky coefficients approximates the original system well even for chaos and the complicated nonequilibrium dynamics in its vicinity. In Figs.~1(b), 1(c), 2(b), 2(c), and 2(d), the moment ODEs \eqref{eq:moment ODE} are truncated to the first $100$ equations, and in the Langevin simulations \eqref{eq:phase oscillators} in Figs.~1(c) and 2(b), the number of oscillators is taken to be $100000$. 

In this paper, we have shown, both theoretically and numerically, that the low-dimensional reduction based on OPUC theory provides a highly accurate approximation to the dynamics of populations of globally coupled phase oscillators with multiharmonic coupling. This theory gives a low-dimensional reduction to a dynamical system on a finite-dimensional manifold represented by Bernstein–Szegő measures, and substantially generalizes the low-dimensional reduction based on the OA ansatz to populations of phase oscillators with arbitrary multiharmonic coupling. Unlike the conventional OA ansatz, however, the reduction is approximate in the case of multiharmonic coupling. This low-dimensional reduction may provide a powerful theoretical foundation for understanding the effects of multiharmonic coupling on collective behavior. Indeed, the numerical simulations presented in this paper confirm that the low-dimensional reduction captures complex nonequilibrium dynamics such as collective chaos and breathing phenomena well, where multiharmonic coupling plays an important role.

Furthermore, the theory developed here is also expected to be highly useful in studies of higher-order interactions in populations of coupled oscillators. This is because phase-oscillator models with higher-order interactions are known to arise through higher-order phase reduction \cite{Leon2019}, and the resulting models have multiharmonic coupling in many cases. Although we have focused here on the case of Kuramoto–Sakaguchi-type coupling, it would be interesting to investigate whether this theory can also be applied to models with more general interactions, including the Winfree model \cite{WINFREE1967,Gallego2017} and the theta-neuron model \cite{Luke2013}. Finally, the present study has shown that OPUC theory can play a significant role in coupled oscillator systems. Further theoretical developments concerning the connection between OPUC theory and coupled oscillator systems are highly anticipated.

\par\vspace{\baselineskip}

\begin{acknowledgments}
The author thanks Hiroshi Kori for valuable discussions. This study was supported by the WINGS-FMSP program at the University of Tokyo.
\end{acknowledgments}

\bibliography{ref}

\beginsupplement

\begin{center}
{\large\bfseries Supplemental Material }\\
\end{center}

\section{Derivation of the ODE for the Verblunsky coefficients}
We now derive the ODE for the Verblunsky coefficients. The monic orthogonal polynomials and the Verblunsky coefficients can be expressed in terms of the moments by the following Heine formulas [1]:
\begin{equation}
\Phi_{n}(z) =\frac{1}{D_{n-1}}
\begin{vmatrix}
Z_0   & Z_1& \cdots &Z_{n-1} & 1 \\
\overline{Z_1}   & Z_0              &  \cdots      & Z_{n-2}&z \\
\vdots& \vdots           &        & \vdots& \vdots\\
\overline{Z_n}    & \overline{Z_{n-1}} & \cdots & \overline{Z_1}&z^{n}
\end{vmatrix}
\end{equation}
\begin{equation}
\|\Phi_n\|^2 = \prod_{j=0}^{n-1} (1 - |\alpha_j|^2) = \frac{D_n}{D_{n-1}}
\end{equation}
\begin{equation}
\alpha_n = (-1)^n\frac{\hat{D}_n}{D_{n}} .
\end{equation}
Here, $D_n$ is the determinant $D_n=\det(T_n)$ of the Toeplitz matrix $T_n$ defined by
\begin{equation}
T_n = (Z_{-j+k})_{0 \le j,k \le n}
= \begin{pmatrix}
Z_0   & Z_1     & \cdots & Z_n \\
Z_{-1}& Z_0     & \cdots & Z_{n-1} \\
\vdots& \vdots  &        & \vdots \\
Z_{-n}& Z_{-n+1}& \cdots & Z_0
\end{pmatrix}
\end{equation}
Note that, as a property of Toeplitz matrices, $D_n > 0$ holds whenever $\rho(\theta,t)d\theta$ is a nontrivial measure. The quantity $\hat{D}_n$ is the determinant $\hat{D}_n = \det(\hat{T}_n)$ of the shifted Toeplitz matrix $\hat{T}_n = (Z_{-j+k+1})_{0 \le j,k \le n}$.

Now, we consider the time derivative of both sides of Eq.~(S3). Using Jacobi's formula for the derivative of a determinant, we obtain
\begin{equation}
\dot{\alpha}_n =
\frac{(-1)^n \operatorname{tr}\big(\operatorname{adj}(\hat{T}_n)\tfrac{d\hat{T}_n}{dt}\big)
      - \alpha_n \operatorname{tr}\big(\operatorname{adj}(T_n)\tfrac{dT_n}{dt}\big)}
     {D_n}\notag.
\end{equation}
Substituting the ODE~(3) for the moments, $\dot{Z}_n =  - \gamma|n| Z_n-in\sum_{l=-L}^{L} h_l Z_{n-l}$ (where \(h_0=\omega\) and \(h_{-l}=\overline{h_l}\)), we obtain
\begin{align}
\dot{\alpha}_n &=
- \gamma \frac{\operatorname{tr}\bigg((-1)^n\operatorname{adj}
(\hat{T}_n)\big(|-j+k+1| Z_{-j+k+1}\big)_{0 \le j,k \le n}- \alpha_n\operatorname{adj}(T_n)\big(|-j+k| Z_{-j+k}\big)_{0 \le j,k \le n}
\bigg)}{D_n} \notag\\
&
- i \sum_{l=-L}^{L} h_l
\frac{\operatorname{tr}\bigg((-1)^n\operatorname{adj}
(\hat{T}_n)\big((-j+k+1) Z_{-j+k+1-l}\big)_{0 \le j,k \le n}-\alpha_n\operatorname{adj}(T_n) \big((-j+k) Z_{-j+k-l}\big)_{0 \le j,k \le n}
\bigg)}{D_n}\notag
\end{align}
 We denote by $R_n$ the trace corresponding to the noise term, and by $S_n^{(l)}$ the trace corresponding to the $l$-th deterministic term, and write
\begin{equation}
\dot{\alpha}_n = - \gamma R_n - i \sum_{l=-L}^{L} h_l S_n^{(l)}
\end{equation}
Furthermore, if $\alpha_n \neq 0$, then $\hat{T}_n$ is nonsingular by the form of the Heine formula (and $T_n$ is always nonsingular when the measure is nontrivial), and therefore these expressions can be rewritten using the inverse matrices as follows:
\begin{eqnarray}
R_n =\alpha_n \operatorname{tr}\bigg(
\hat{T}_n^{-1}\big(|-j+k+1| Z_{-j+k+1}\big)_{0 \le j,k \le n}- T_n^{-1}\big(|-j+k| Z_{-j+k}\big)_{0 \le j,k \le n}
\bigg)
\end{eqnarray}
\begin{eqnarray}
S_n^{(l)} =
\alpha_n\operatorname{tr}\bigg(
\hat{T}_n^{-1}\big((-j+k+1) Z_{-j+k+1-l}\big)_{0 \le j,k \le n}- T_n^{-1}\big((-j+k) Z_{-j+k-l}\big)_{0 \le j,k \le n}
\bigg)
\end{eqnarray}

First, we consider (S7). Using the definition of the moments, we obtain

\begin{align}
&\operatorname{tr}\bigg(
T_n^{-1}\big((-j+k) Z_{-j+k-l}\big)_{0 \le j,k \le n}
\bigg)\notag\\
&= \sum_{j=0}^{n}\sum_{k=0}^{n} (-j+k) (T_n^{-1})_{j,k} Z_{-j+k-l}\notag \\
&= \int_0^{2\pi} i e^{i l \theta}
\frac{\partial}{\partial \theta}
\bigg\{
\sum_{j=0}^{n}\sum_{k=0}^{n}
(T_n^{-1})_{j,k} e^{i (j-k)\theta}
\bigg\} \,\rho(\theta,t)\,d\theta \notag\\
&= \int_0^{2\pi} i e^{i l \theta}
\frac{\partial}{\partial \theta}
\big\{
\bm{v}_n(\theta)^{*} T_n^{-1} \bm{v}_n(\theta)
\big\} \,\rho(\theta,t)\,d\theta\notag
\end{align}

where $\bm{v}_n(\theta) = (1, e^{i\theta}, \ldots, e^{i n\theta})^T$, and $\bm{v}_n(\theta)^{*}$ denotes its conjugate transpose.

Here, using the ABC theorem for the Christoffel--Darboux kernel [2] $K_n(z,\zeta) = \sum_{j=0}^{n} \frac{\overline{\Phi_j(\zeta)}\Phi_j(z)}{\|\Phi_j\|^2} $, namely $T_n^{-1} = k^{(n)}$ (where $k^{(n)}$ is the matrix consisting of the coefficients $k_{j,k}^{(n)}$ appearing in the expansion $K_n(z,\zeta) = \sum_{j=0}^{n}\sum_{k=0}^{n} k_{j,k}^{(n)} \overline{\zeta^j} z^k$), we obtain
\begin{align}
&\operatorname{tr}\bigg(
T_n^{-1}\big((-j+k) Z_{-j+k-l}\big)_{0 \le j,k \le n}
\bigg)\notag\\
&= \int_0^{2\pi} i e^{i l \theta}
\frac{\partial}{\partial \theta}
\big\{
\bm{v}_n(\theta)^{*} T_n^{-1} \bm{v}_n(\theta)
\big\} \, \rho(\theta,t)\,d\theta \notag\\
&= \int_0^{2\pi} i e^{i l \theta}
\frac{\partial}{\partial \theta}
\big\{
K_n(e^{i\theta}, e^{i\theta})
\big\} \, \rho(\theta,t)\,d\theta
\end{align}
Next, letting $A = (Z_1,\ldots,Z_n)$ and $D = (Z_n,\ldots,Z_1)^T$, we consider the following block matrix:
\begin{eqnarray}
\hat{T}_n =
\begin{pmatrix}
A      & Z_{n+1} \\
T_{n-1}& D\notag
\end{pmatrix}.
\end{eqnarray}
Then,
\begin{align}
&\operatorname{tr}\bigg(\hat{T}_n^{-1}\big(
(-j+k+1) Z_{-j+k+1-l}
\big)_{0 \le j,k \le n}\bigg)\notag\\
&= \int_0^{2\pi} i e^{i l \theta}
\frac{\partial}{\partial \theta}
\big\{
\bm{v}_n(\theta)^{*} \hat{T}_n^{-1} \bm{v}_n(\theta)e^{-i\theta}
\big\} \, \rho(\theta,t)\,d\theta \notag\\
&= \int_0^{2\pi} i e^{i l \theta}
\frac{\partial}{\partial \theta}
\bigg\{
\begin{pmatrix}
\bm{v}_{n-1}(\theta)^{*} & e^{-i n\theta}
\end{pmatrix}
\begin{pmatrix}
A      & Z_{n+1} \\
T_{n-1}& D
\end{pmatrix}^{-1}
\begin{pmatrix}
1 \\
e^{i\theta}\bm{v}_{n-1}(\theta)
\end{pmatrix}e^{-i\theta}
\bigg\} \, \rho(\theta,t)\,d\theta\notag\\
&= \int_0^{2\pi} i e^{i l \theta}
\frac{\partial}{\partial \theta}
\bigg\{K_{n-1}(e^{i\theta},e^{i\theta})+e^{-i\theta}\frac{(1-e^{in\theta}\bm{v}_{n-1}(\theta)^{*}T_{n-1}^{-1}D)(1-e^{i\theta} A T_{n-1}^{-1}\bm{v}_{n-1}(\theta))}{e^{in\theta}(Z_{n+1}-AT_{n-1}^{-1}D)}\big\} \, \rho(\theta,t)\,d\theta\notag
\end{align}
Here, by the determinant formula for block matrices,
\[
\det \hat{T}_n = (-1)^n\det T_{n-1}\,(Z_{n+1} - A T_{n-1}^{-1} D),
\]
and hence
\[
Z_{n+1} - A T_{n-1}^{-1} D =  \alpha_n \|\Phi_n\|^2.
\]
Furthermore, since
\[
1 - e^{i n\theta} \bm{v}_{n-1}(\theta)^{*} T_{n-1}^{-1} D = \Phi_n^*(e^{i\theta}),
\qquad
\frac{1 - e^{i\theta} A T_{n-1}^{-1} \bm{v}_{n-1}(\theta)}{e^{i n\theta}}
= \overline{\Phi_n(e^{i\theta})},
\]
it follows that
\begin{align}
&\operatorname{tr}\bigg(\hat{T}_n^{-1}\big(
(-j+k+1) Z_{-j+k+1-l}
\big)_{0 \le j,k \le n}\bigg) \notag\\
&=\int_0^{2\pi} i e^{i l \theta}
\frac{\partial}{\partial \theta}
\bigg\{
K_{n-1}(e^{i\theta},e^{i\theta})
+ e^{-i\theta}
\frac{\Phi_n^*(e^{i\theta})\,\overline{\Phi_n(e^{i\theta})}}{ \alpha_n \|\Phi_n\|^2}
\bigg\} \,\rho(\theta,t)\,d\theta 
\end{align}
From (S8) and (S9), we obtain the following.
\begin{align}
S_n^{(l)}
&= \alpha_n \int_0^{2\pi} i e^{i l \theta}
\frac{\partial}{\partial \theta}
\bigg\{
K_{n-1}(e^{i\theta},e^{i\theta})
- K_n(e^{i\theta},e^{i\theta})
+ e^{-i\theta}
\frac{\Phi_n^*(e^{i\theta})\,\overline{\Phi_n(e^{i\theta})}}{ \alpha_n \|\Phi_n\|^2}
\bigg\} \,\rho(\theta,t)\,d\theta \notag\\
&= \frac{1}{\|\Phi_n\|^2}
\int_0^{2\pi} i e^{i l \theta}
\frac{\partial}{\partial \theta}
\bigg\{
e^{-i\theta} \Phi_n^*(e^{i\theta})\,\overline{\Phi_n(e^{i\theta})}
- \alpha_n |\Phi_n(e^{i\theta})|^2
\bigg\} \,\rho(\theta,t)\,d\theta \notag\\
&= \frac{1}{\|\Phi_n\|^2}
\int_0^{2\pi} i e^{i l \theta}
\frac{\partial}{\partial \theta}
\bigg\{
\Phi_n^*(e^{i  \theta})\,\overline{\Phi_{n+1}(e^{i \theta})}
\bigg\} \, \rho(\theta,t)\,d\theta 
\end{align}
where we used $K_n(e^{i\theta},e^{i\theta}) - K_{n-1}(e^{i\theta},e^{i\theta}) = \frac{|\Phi_n(e^{i\theta})|^2}{\|\Phi_n\|^2}$ and the Szeg\H{o} recurrence.

Up to this point, we have carried out the calculation under the assumption that $\alpha_n \ne 0$. We now consider the case $\alpha_n = 0$. In this case, $\hat{T}_n$ is not invertible, and since it contains $T_{n-1}$ as a submatrix, its rank is seen to be $n$. Therefore, $\operatorname{adj}(\hat{T}_n)$ has rank $1$. Hence, there exists a scalar $d$ such that, using vectors $\bm{x}$ and $\bm{y}^T$ satisfying $\hat{T}_n \bm{x} = \bm{0}$ and $\bm{y}^T \hat{T}_n = \bm{0}$, we can write
\begin{equation}
\operatorname{adj}(\hat{T}_n) = d \bm{x} \bm{y}^T
\end{equation}
Here, using the coefficients of $\Phi_n^{*}(z)=\sum_{m=0}^{n}d_{m}^{(n)} z^m$, we define
$
\bm{x} = (d_n^{(n)}, \ldots, d_0^{(n)})^T,
\bm{y} = (d_0^{(n)}, \ldots, d_n^{(n)})^T
$. Then, by the Szeg\H{o} recurrences, when $\alpha_n = 0$ we have
$
\Phi_{n+1}(z) = z \Phi_n(z), \Phi_{n+1}^*(z) = \Phi_n^*(z)
$
and hence
\begin{align}
(\hat{T}_n \bm{x})_j
&= \sum_{k=0}^{n} d_{n-k}^{(n)} Z_{-j+k+1} \notag\\
&= \big\langle z \sum_{k=0}^{n} \overline{d_{n-k}^{(n)}} z^k, z^j \big\rangle \notag\\
&= \langle \Phi_{n+1}(z), z^j \rangle = 0 \qquad (j = 0,\ldots,n)\notag
\end{align}
\begin{align}
(\bm{y}^T \hat{T}_n)_k
&= \sum_{j=0}^{n} d_j^{(n)} Z_{-j+k+1} \notag\\
&= \big\langle z^{k+1}, \sum_{j=0}^{n} d_j^{(n)} z^j \big\rangle \notag\\
&= \langle z^{k+1}, \Phi^*_{n+1}(z) \rangle = 0 \qquad (k = 0,\ldots,n)\notag
\end{align}
which shows that this choice of $\bm{x}$ and $\bm{y}^T$ is appropriate.

Furthermore, since the $(n+1,1)$ entry of the cofactor matrix (that is, $\operatorname{adj}(\hat{T}_n)$) is $(-1)^{n+2} D_{n-1}$, we have $d = (-1)^{n+2} D_{n-1}$. Therefore, when $\alpha_n = 0$, $S_n^{(l)}$ is given by
\begin{align}
S_n^{(l)}&= \frac{(-1)^n}{D_n}\operatorname{tr}\bigg(
\operatorname{adj}(\hat{T}_n)\big((-j+k+1) Z_{-j+k+1-l}\big)_{0 \le j,k \le n}
\bigg) \notag\\
&= \frac{(-1)^n}{D_n}\operatorname{tr}\bigg(
d\,\bm{x}\bm{y}^{T}\big((-j+k+1) Z_{-j+k+1-l}\big)_{0 \le j,k \le n}
\bigg) \notag\\
&= \frac{D_{n-1}}{D_n}
\sum_{j=0}^{n} \sum_{k=0}^{n} (-j+k+1)
d_j^{(n)}\,d_{n-k}^{(n)}\, Z_{-j+k+1-l} \notag\\
&= \frac{1}{\|\Phi_n\|^2}
\int_0^{2\pi} i e^{i l \theta}
\frac{\partial}{\partial \theta}
\bigg\{
\Phi_n^*(e^{i  \theta})\,\overline{\Phi_{n+1}(e^{i \theta})}
\bigg\} \, \rho(\theta,t)\,d\theta \notag
\end{align}
and thus Eq. (S10) is  obtained. Therefore, $S_n^{(l)}$ always takes the form given in (S10).

The noise term $R_n$ can be derived in exactly the same way by replacing $\frac{\partial}{\partial \theta}$ with $|\frac{\partial}{\partial \theta}|$, yielding
\begin{align}
R_n
&= -\frac{1}{\|\Phi_n\|^2}
\int_0^{2\pi} 
|\frac{\partial}{\partial \theta}|
\bigg\{
e^{-i\theta} \Phi_n^*(e^{i \theta})\,\overline{\Phi_n(e^{i \theta})}
- \alpha_n |\Phi_n(e^{i \theta})|^2
\bigg\} \, \rho(\theta,t)\,d\theta \notag\\
&= -\frac{1}{\|\Phi_n\|^2}
\int_0^{2\pi} 
|\frac{\partial}{\partial \theta}|
\bigg\{
 \Phi_n^*(e^{i \theta})\,\overline{\Phi_{n+1}(e^{i \theta})}\bigg\}\, \rho(\theta,t)\,d\theta
\end{align}
Thus, we finally obtain
\begin{align}
\dot{\alpha}_n &= - \gamma R_n - i \sum_{l=-L}^{L} h_l S_n^{(l)} \notag\\
&= \frac{\gamma}{\|\Phi_n\|^2}
\int_0^{2\pi} 
\left|\frac{\partial}{\partial \theta}\right|
\bigg\{
 \Phi_n^*(e^{i \theta})\overline{\Phi_{n+1}(e^{i \theta})}\bigg\} \rho(\theta,t)d\theta+\frac{1}{\|\Phi_n\|^2}
\int_0^{2\pi}  h(\theta,t)
\frac{\partial}{\partial \theta}
\bigg\{
 \Phi_n^*(e^{i  \theta})\overline{\Phi_{n+1}(e^{i \theta})}\bigg\} \rho(\theta,t)d\theta
\end{align}

Moreover, by differentiating the Heine formula for the monic orthogonal polynomials (S1) with respect to time, one obtains in exactly the same way the following ODE for the monic orthogonal polynomials:
\begin{align}
\dot{\Phi}_n(z) &=- \gamma
\int_0^{2\pi} 
|\frac{\partial}{\partial \theta}|
\bigg\{
 \Phi_n(e^{i  \theta})\,K_{n-1}(z, e^{i\theta})\bigg\}\, \rho(\theta,t)d\theta
-
\int_0^{2\pi}  h(\theta,t)
\frac{\partial}{\partial \theta}
\bigg\{
 \Phi_n(e^{i  \theta})\,K_{n-1}(z, e^{i\theta}) \bigg\}\, \rho(\theta,t)d\theta
\end{align}
Setting $z=0$ in this ODE, we obtain the ODE for the Verblunsky coefficients.

\section{Results obtained from the ODE for the Verblunsky coefficients}
In what follows, using the ODE for the Verblunsky coefficients derived above, we observe that for $n\ge 2L-1$, $\alpha_n$ is not directly driven by the first $L$ Verblunsky coefficients. Let $\dot{\alpha}_n|_{\rho_N}$ denote the flow of $\dot{\alpha}_n$ computed on the $N$th Bernstein--Szeg\H{o} measure. Below, we consider $\dot{\alpha}_n|_{\rho_L}$ for $n\ge 2L-1$:
\begin{align}
\dot{\alpha}_n|_{\rho_L} &= \frac{\gamma}{\|\Phi_L\|^2}
\int_0^{2\pi} 
|\frac{\partial}{\partial \theta}|
\bigg\{
 \Phi_L^*(e^{i \theta})^2e^{-i(n+1) \theta}\bigg\} \rho_L(\theta,t)\,d\theta+\frac{1}{\|\Phi_L\|^2}
\int_0^{2\pi}  h(\theta,t)
\frac{\partial}{\partial \theta}
\bigg\{
 \Phi_L^*(e^{i \theta})^2e^{-i(n+1) \theta}\bigg\}\rho_L(\theta,t)\,d\theta \notag\\
 &=\big\langle (i\gamma+h(z,t))(-2iz\overline{\frac{d\Phi_L^*(z)}{dz}}+i(n+1)\overline{\Phi_L^*(z)})z^{n+1},\Phi_L^*(z) \big\rangle_{\rho_L} 
\end{align}
Here, since the expression inside $|\frac{\partial}{\partial \theta}|$ consists entirely of negative-frequency modes, we have rewritten it as $|\frac{\partial}{\partial \theta}|=-i\frac{\partial}{\partial \theta}$. Also, the reversed polynomial of the monic orthogonal polynomial appearing above is taken with respect to the degree-$L$ Bernstein--Szeg\H{o} measure $\rho_L(\theta,t)\,d\theta$. The left-hand factor in the inner product consists only of terms of degree at least $1$. For the Bernstein--Szeg\H{o} measure, the reversed polynomial is orthogonal to all terms of degree at least $1$, and therefore $\dot{\alpha}_n|_{\rho_L}=0$ holds for $n\ge 2L-1$.

Recalling the definition of the Bernstein--Szeg\H{o} measure, the fact that $\dot{\alpha}_n|_{\rho_L}=0$ for $n\ge 2L-1$ means that the flow of $\dot{\alpha}_n$ contains no terms depending only on $\alpha_j\ (0\le j\le L-1)$. As a first consequence, when $L=1$, the flow of $\dot{\alpha}_n\ (n\ge 1)$ contains no term depending only on $\alpha_0$, so $\alpha_n=0\ (n\ge 1)$ forms an invariant manifold. This corresponds to the Ott--Antonsen ansatz. On the other hand, for $L\ge 2$, $\alpha_n=0\ (n\ge L)$ does not form an invariant manifold; however, we do find that $\alpha_n\ (n\ge 2L-1)$ is not directly driven by the first $L$ Verblunsky coefficients, which are nonzero for the uniformly rotating solution. By computing $\dot{\alpha}_n|_{\rho_N}$ for general $N$, one can also verify that for $L\ge2$, no Bernstein--Szeg\H{o} measure forms an invariant manifold.

\section{Derivation of Eq. (13)}
Next, using this, we consider the order of $\alpha_n$ for $n\ge L$. Let
$T_n^{qst}=(Z^{qst}_{-j+k})_{0 \le j,k \le n}$,
$\hat{T}_n^{qst}=(Z^{qst}_{-j+k+1})_{0 \le j,k \le n}$,
$U_n=(u_{-j+k})_{0 \le j,k \le n}$, and $\hat{U}_n=(u_{-j+k+1})_{0 \le j,k \le n}$.
Expanding (S3) up to $O(\varepsilon)$ using Jacobi's formula, we obtain
\begin{eqnarray}
\alpha_n = \varepsilon(-1)^n \frac{\operatorname{tr}\big(\operatorname{adj}(\hat{T}_n^{qst})\hat{U}_n\big)}{\det T_n^{qst}} +O(\varepsilon^2)
\end{eqnarray}
Here, since $\det \hat{T}_n^{qst}=0$, we see that no $O(1)$ term remains. Since $\operatorname{adj}(\hat{T}_n^{qst})$ has rank $1$, it can be expanded in the same way as in (S11). In the present case, since we are considering a degree-$L$ Bernstein--Szeg\H{o} measure, the coefficients in
$\Phi_n^{*}(z)=\sum_{m=0}^{n}d_{m}^{(n)} z^m$
satisfy
$d_j^{(n)}=d_j\,(0\le j\le L)$ and $d_j^{(n)}=0\,(L+1\le j \le n)$.
Therefore, if we choose
$
\bm{x} = (0, \ldots,0, d_L,\ldots,d_0)^T, 
\bm{y} = (d_0, \ldots,d_L, 0,\ldots,0)^T,
$
then we obtain
\begin{eqnarray}
\alpha_n &=& \varepsilon\frac{(-1)^n}{D_n^{qst}}\operatorname{tr}\big(
d\,\bm{x}\bm{y}^{T}\hat{U}_n
\big)+O(\varepsilon^2) \\
&=& \frac{\varepsilon}{\|\Phi_L^{*\,qst}\|^2}
\sum_{j=0}^{L} \sum_{k=0}^{L} 
d_j\,d_k\, u_{(n+1)-(j+k)}+O(\varepsilon^2)
\end{eqnarray} 
where we used $d = (-1)^{n+2} \det T_{n-1}^{qst}$.

\suppsection{References}
[1] B. Simon, Orthogonal Polynomials on the Unit Circle. Part 1: Classical Theory, Colloquium Publications,
Vol. 54 (American Mathematical Society, Providence, RI,
2005).

[2] B. Simon, The Christoffel–Darboux kernel, in Perspectives in Partial Differential Equations, Harmonic Analysis and Applications: A Volume in Honor of Vladimir G.
Maz’ya’s 70th Birthday, Proceedings of Symposia in Pure
Mathematics, Vol. 79, edited by D. Mitrea and M. Mitrea
(American Mathematical Society, Providence, RI, 2008)
pp. 295–335.

\end{document}